\documentclass[reprint,amsmath,amssymb,aps,prb]{revtex4-2}
\usepackage{graphicx}
\usepackage{color}
\usepackage{dcolumn}
\usepackage{latexsym}

\usepackage[normalem]{ulem}
\usepackage{hyperref,amssymb}
\usepackage{url}
\usepackage{color,soul}
\usepackage{graphicx}
\usepackage{verbatim}
\usepackage{multirow}
\usepackage{amsmath}
\newcommand{\beq}{\begin{eqnarray}}
\newcommand{\eeq}{\end{eqnarray}}
\usepackage{mathrsfs}
\usepackage{float,soul}
\usepackage[dvipsnames]{xcolor}
\usepackage{mathtools}
\usepackage{slashed}
\usepackage{physics}
\usepackage{graphicx}
\usepackage{epstopdf}
\usepackage{subfigure}
\usepackage[font=small]{caption}
\usepackage{hyperref}
\usepackage{bbold}
\usepackage{wasysym}
\usepackage{feynmp}
\usepackage{hyperref}
\hypersetup{colorlinks}
\usepackage{ragged2e}
\usepackage{siunitx}
\usepackage[utf8]{inputenc}

\def\red{\color{red}}

\begin{document}

\title{Coherently synchronized oscillations in many-body localization}
\author{Zi-Jian Li}
\thanks{These authors contributed equally to this work}
\author{Yi-Ting Tu}
\thanks{These authors contributed equally to this work}
\author{Sankar Das Sarma}
\affiliation{Condensed Matter Theory Center and Joint Quantum Institute, Department of Physics, University of Maryland, College Park, Maryland 20742, USA}

\begin{abstract}
{We find an unexpected phenomenon of coherently synchronized oscillations in a mirror-symmetric many-body localized system. A synchronization transition of the spin oscillations is found by changing the spin-spin interactions. To understand this phenomenon, an effective Ising model based on local integrals of motion is proposed. We find that the synchronization transition can be understood as a paramagnetic-to-ferromagnetic Ising transition. Based on the Ising model, we theoretically estimate the synchronized frequencies and the synchronization transition points, which agree well with numerical results.}
\end{abstract}
\maketitle

Synchronization is a collective phenomenon that has been widely observed in various fields, including physics, chemistry, and biology~\cite{pikovsky2001synchronization}. In the paradigmatic Kuramoto model of coupled self-sustained oscillators~\cite{kuramoto1984chemical,acebron2005kuromoto}, dissipation and interaction both play important roles in synchronization, where dissipation is necessary to stabilize the oscillators, known as limit-cycle oscillators~\cite{van1926on}, and interaction is required to synchronize independent oscillations. 

Although it is commonly believed that strict synchronization cannot appear in conservative systems due to the lack of limit cycles, synchronized behaviors occur widely in both quantum and classical conservative systems~\cite{hampton1999measure,wang2003measure,qiu2014measure}. In conservative systems with many degrees of freedom, synchronized behaviors are noisy due to ergodicity, where any trajectories with different initial conditions eventually explore the whole phase space and cannot be distinguished from each other. 
However, ergodicity may completely break down in quantum many-body systems through eigenstate localizations, known as \emph{many-body localization} (MBL)~\cite{fleishman1980interactions,basko2006metal,oganesyan2007localization,abanin2019manybody,alet2018many,sierant2025many}, where one can expect a persistent non-ergodic motion of particles under certain initial conditions. Since an MBL system can be described as adding interaction to an Anderson localized system, which consists of independent localized orbitals, it is natural to ask whether there can be oscillatory motions arising from these orbitals, which possibly can be synchronized when interaction is introduced.

In this Letter, we theoretically and numerically discover synchronized behaviors in a coupled MBL system. We show that the oscillations of spins, whose initial configuration is a product state, can be synchronized due to nearest-neighbor spin-spin interactions. More specifically, a synchronization transition is found by changing the strength of the interactions. We introduce a theoretical model based on the local integrals of motion (LIOMs), where the synchronization transitions can be understood as a paramagnetic-ferromagnetic transition of an effective Ising model. The validity of the theory is examined with numerical calculations on synchronized oscillation frequencies and quantum correlations. Finally, we discuss the robustness of the oscillations to experimental imperfections and the similarity of the phenomenon to classical synchronization in conservative systems. 

\begin{figure}
    \centering
    \includegraphics[width=\linewidth]{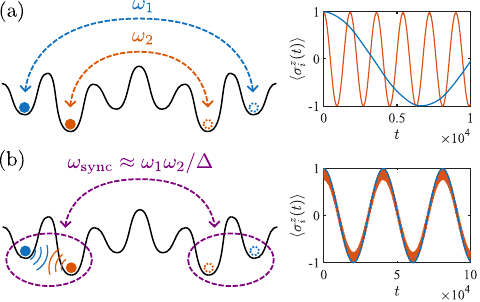}
    \caption{\justifying{Coherently synchronized oscillations in a mirror-symmetric MBL system. (a) Unsynchronized oscillations in the non-interacting system with frequencies $\omega_1$ and $\omega_2$ (drawn color-wise). (b) Synchronized oscillations in the interacting system with frequency $\omega_\text{sync}$. A small interaction $\Delta=0.01$ is used so that the approximation $\omega_1\omega_2/\Delta$ holds. For both figures, we consider a single disorder realization at $W=8,L=6$}}
    \label{fig1}
\end{figure}

\textit{Setup.} -- We consider a disordered spin-1/2 XXZ chain satisfying a global mirror symmetry
\begin{equation}
    H = H_0 + P H_0 P + H_{\rm cp}
    \label{eq:model}
\end{equation}
where
\begin{equation}
\begin{aligned}
    & H_0 = \frac{1}{4}\sum_{i=1}^{\frac L2-1} \left(\sigma_i^x \sigma_{i+1}^x + \sigma_i^y \sigma_{i+1}^y + \Delta \sigma_i^z \sigma_{i+1}^z \right) + \frac{1}{2}\sum_{i=1}^{\frac L2} h_i \sigma_i^z, \\
    & H_{\rm cp} = \frac{1}{4} \left(\sigma_{\frac L2}^x \sigma_{\frac L2+1}^x + \sigma_{\frac L2}^y \sigma_{\frac L2+1}^y + \Delta \sigma_{\frac L2}^z \sigma_{\frac L2+1}^z\right), \\
\end{aligned}
\end{equation}
where $\sigma^{x,y,z}$ are the Pauli operators, $\Delta$ is the interaction strength between nearby spins, the onsite disorder $h_i$ is randomly sampled from the interval $[-W,W]$, and the mirror symmetry operator satisfies $P \sigma_i^\alpha P = \sigma_{\bar i}^\alpha, \bar i = L-i+1$.
Later, we will treat $H_{\rm cp}$ and $\Delta$ as perturbations in the theory (except when we discuss the strongly interacting regime).
The fermionic representation can be introduced according to the Jordan-Wigner transformation. We will switch between these two different representations for convenience.  In the non-interacting case ($\Delta=0$), the half of the spin chain $H_0$ is Anderson localized. With non-zero interactions, $H_0$ shows an ergodic-to-MBL transition separated by a critical disorder $W_c$. The numerically estimated critical disorder for $L\sim 16$ and $\Delta=1$ is around $W_c\sim 3.5$\cite{pal2010manybody,de2013ergodicity,luitz2015manybody}. Our model can be realized in different MBL experimental platforms\cite{schreiber2015observation,choi2016exploring,kai2018emulating}. Notice that a similar MBL system has been realized in Ref.~\cite{bordia2016coupling}.

To prepare a single oscillator, we let the initial state be a one-particle state on site $j$, and evaluate the time evolution of the amplitude on each site $c_{k}(t) = \langle k|\psi(t)\rangle$. 
Since the single-particle eigenstates of $H_0$ are exponentially localized around charge centers $\langle j|\phi_i \rangle \sim e^{-|i-j|/\xi}$, to first-order perturbation, the eigenstates of $H$ are $\frac{1}{\sqrt 2}(|\phi_i\rangle \pm |\phi_{\bar i}\rangle)$, with eigenvalues $E_i \pm \omega_i$. $\xi$ is the localization length and $\omega_i \sim e^{-(L-2j)/\xi}$ comes from the coupling term $H_\text{cp}$. Note that here $\Delta=0$. Thus, we have
\begin{equation}
\begin{aligned}
     c_k(t) = \sum_{i=1}^{L/2}  \langle \phi_i|j\rangle e^{i E_i t} [\langle k|\phi_i\rangle \cos(\omega_i t) + \langle k|\phi_{\bar i}\rangle i\sin(\omega_i t)]
\end{aligned}
\end{equation}
For sufficiently small localization length $\xi$, we have $\langle j|\phi_i \rangle \approx \delta_{i,j}$, where $\delta_{i,j}$ is the Kronecker delta function. Therefore
\begin{equation}
    |c_k(t)|^2 \approx \delta_{k,j} \cos^2(\omega_j t) + \delta_{k,\bar j} \sin^2(\omega_j t)
\end{equation}
This means a particle will approximately oscillate back and forth between site $j$ and its mirror site $\bar j$, analogous to an oscillator. 


\textit{Two particles.} -- The physical picture of synchronization in our system can be illustrated with two particle oscillations. Let the initial state be the product of non-interacting eigenstates prepared at one side of the chain $|\psi_0\rangle = c_i^\dagger c_j^\dagger |0\rangle$, where $c_i(c_i^\dagger)$ annihilates (creates) a non-interacting localized orbital around site $i$. 
Without $\Delta$ and $H_{\rm cp}$, the initial state lies in a 4-fold degenerate subspace, expanded by mirror reflections of the eigenstates $\left\{c_i^\dagger c_j^\dagger |0\rangle, c_i^\dagger c_{\bar j}^\dagger |0\rangle,c_{\bar i}^\dagger c_j^\dagger |0\rangle,c_{\bar i}^\dagger c_{\bar j}^\dagger |0\rangle\right\}$. 
After adding the interaction, this 4-fold subspace becomes two 2-fold degenerate subspaces $\left\{c_i^\dagger c_j^\dagger |0\rangle, c_{\bar i}^\dagger c_{\bar j}^\dagger |0\rangle\right\}$ and $\left\{c_i^\dagger c_{\bar j}^\dagger |0\rangle,c_{\bar i}^\dagger c_j^\dagger |0\rangle\right\}$, whose energy difference is approximately $\delta E\sim \Delta |e^{-|i-j|/\xi} - e^{-|i-\bar j|/\xi}|$. Now we turn on $H_{\rm cp}$. 
In the non-interacting case, two particles oscillate independently with different frequencies $\omega_i$ and $\omega_j$. 
In the interacting case, if $\delta E \gg \omega_i+\omega_j$, the tunneling amplitude from one subspace to the other is suppressed. Thus, the initial states can only evolve within one subspace. Phenomenologically, two particles are bonded together and oscillate in a synchronized way (Fig.~\ref{fig1}). A related situation is shown in Ref.~\cite{kleinherbers2021synchronized} in a different setup. 

Next, we introduce our main theoretical picture of synchronization for the genuine many-body scenario based on the above idea of perturbation, with the complication that the degenerate subspaces are now themselves many-body systems.

\begin{figure}
    \centering
    \includegraphics[width=\linewidth]{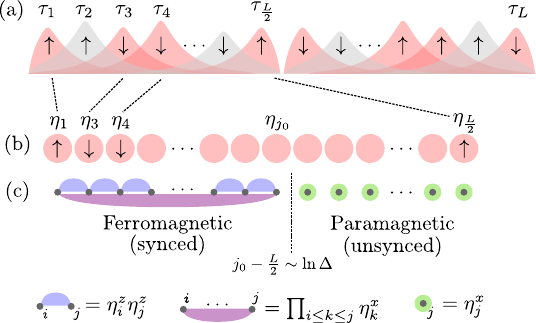}
    \caption{\justifying {Illustration of the theory of synchronization in MBL. (a) The LIOMs of the uncoupled chain, along with an example configuration, where red (gray) indicates active (inactive) sites. (b) The effective $\eta$ Ising chain formed by the active sites of the left half of the $\tau$ chain. (c) The approximated integrals of motion of the $\eta$ chain.}}
    \label{fig:ising}
\end{figure}
\textit{Theoretical picture.}--
Our theory is based on treating $H_{\rm cp}$ and $\Delta$ as perturbations.
When $H_{\rm cp}$ is turned off, the system becomes two independent localized spin chains.
As we assume Hamiltonian $H_0$ of the left chain to be many-body localized, there exists a complete set of LIOMs $\tau^z_i$, localized around site $i$, with $i=1,\cdots,\frac{L}{2}$.
We can then write the Hamiltonian in terms of such LIOMs as~\cite{Serbyn2013,Ros2015,Chandran2015,Imbrie2017}
\begin{equation}
    H_0=\sum_{i=1}^{L/2} h'_i \tau^z_i
    +\sum_{i,j=1}^{L/2}J'_{ij}\tau^z_i\tau^z_j
    +\sum_{ijk}J'_{ijk}\tau^z_i\tau^z_j\tau^z_k+\cdots,
\end{equation}
In the non-interacting (Anderson localized) case, all couplings between LIOMs vanish, while in the interacting (MBL) case, $J'_{ij\cdots}$ decay exponentially with the farthest distance among $i,j,\ldots$. In addition, there exist operators $\tau^{x}_i$ and $\tau^{y}_i$, also localized around $i$, such that the $\{\tau^{x,y,z}_i\}$ operators satisfy the same algebra as $\{\sigma^{x,y,z}_i\}$.
The mirrored chain on the right can be expanded similarly with $\tau^{x,y,z}_{\bar i}=P\tau^{x,y,z}_iP$.

To couple the two chains, we write $H_{\rm cp}$ in terms of the $\tau$ operators as
\begin{equation}
    H_{\rm cp}=\sum_{r,r'=1}^{L/2} O_{r,r'},\quad ||O_{r,r'}|| \sim e^{-\frac{r+r'}{\xi}}
\end{equation}
where $O_{r,r'}$ is a sum of Pauli strings of $\tau_i$ with $i=\frac{L}{2}-r+1,\ldots,\frac{L}{2}+r'$).

Now we treat $H_{\rm cp}$ as a perturbation~\footnote{
As $\lVert H_{\rm cp} \rVert$ is not small, a more accurate description is to move $O_{r,r'}$ with $r, r' \lesssim \xi$ into the unperturbed Hamiltonian, which reconfigures the LIOMs near $j = \frac{L}{2}$, and to treat only the rest of the sum as a perturbation. However, this is effectively equivalent to treating $H_{\rm cp}$ itself as a perturbation and trusting the results only away from $j = \frac{L}{2}$.
}.
In the non-interacting case, the unperturbed Hamiltonian is $H^{(0)}=\sum_{i=1}^L h_i'\tau_i^z$ with $h'_{\bar i}=h'_i$, whose eigenspaces can be labeled by $a_i:=\langle\frac{\tau^z_i+\tau^z_{\bar i}}{2}\rangle=0,\pm1$ for $i=1,\ldots\frac{L}{2}$.
Calling the sites with $a_i=0$ the \emph{active sites}, we can see that (assuming no accidental degeneracy) the eigenspace is $2^A$-fold degenerate, where $A$ is the number of active sites from site $1$ to $\frac{L}{2}$.
We write this eigenspace as an effective spin chain of length $A$ with $|0\rangle_j$ and $|1\rangle_j$ corresponds to $|01\rangle_{j,\bar j}$ and $|10\rangle_{j,\bar j}$ in the $\tau$ chain, and write the Pauli operators of the effective chain as $\eta^{x,y,z}_j$. A similar mapping has been considered in Ref.~\cite{thomas2018quantum} in a different context.
For notational convenience, we let the site label $j$ in the $\eta$ chain run over all active sites in the left half of the original chain, rather than from $1$ to $A$ (see Fig.~\ref{fig:ising}a--b).
As $\tau_i^z=\pm 1$ in the non-interacting case is equivalent to the $i$th orbital being filled/unfilled, the action of $O_{r,r'}$ is to simply move a particle between orbital $\frac{L}{2}-r+1$ and $\frac{L}{2}+r'$.
This implies that the effective Hamiltonian in each degenerate subspace is, to lowest order,
\begin{equation}
    H_\text{eff}=\sum_j h^\text{eff}_j\eta_j^x,\quad h^\text{eff}_j\sim e^{-\frac{L-2j}{\xi}}.
\end{equation}
That is, it is an Ising model with exponentially decaying transverse field in the purely paramagnetic phase.
If $\xi\lesssim 1$, the physical spin $\sigma^z_i$ has the largest overlap with $\tau^z_i$, which in turn equals to $\eta^z_i$ if $a_i=0$.
So in our dynamical setup, the oscillation frequencies of $\langle\sigma^z_j(t)\rangle$ will be dominated by that of $\langle\eta^z_j(t)\rangle$ on the corresponding $\eta^z$ eigenstate.
That is, $\omega_j\approx |2h^\text{eff}_j|$.

In the presence of weak interaction, the $J'_{ij\cdots}$ terms can be added to the perturbation Hamiltonian.
Keeping only the most important terms across each site and bond, we have
\begin{equation}
    H_\text{eff}=\sum_{\langle ij\rangle}J^\text{eff}_{ij}\eta_i^z\eta_j^z+\sum_j h^\text{eff}_j\eta_j^x,\quad J^\text{eff}_{ij}\sim \Delta e^{-\frac{|i-j|}{\xi}}.
\end{equation}
where $\langle ij\rangle$ runs over neighboring pairs of the $\eta$ chain.
That is, the effect of interaction is essentially to add an Ising coupling term in the $\eta$ chain.
Suppose the distribution of active sites is approximately uniform, then the magnitude of $J^\text{eff}$ is also uniform (and fluctuating) in the $\eta$ chain.
On the other hand, $h^\text{eff}$ is exponentially decaying (and disordered) from the right to the left, which divides the $\eta$ chain into a ferromagnetic (FM) region with $h^\text{eff}<J^\text{eff}$, and a paramagnetic (PM) region with $h^\text{eff}>J^\text{eff}$, separated around a critical site $j_0$
\begin{equation} \label{eq:critical}
    L-2j_0\approx-\xi\ln\Delta+\text{const.}
\end{equation}

Due to the disorder of $h^\text{eff}$ and $J^\text{eff}$~\footnote{
    One may wonder why when all sites are active, $J^\text{eff}$ is always approximately $\Delta/2$, raising the question whether the FM region has enough disorder to be MBL. However, the slight fluctuation around $\Delta/2$ becomes strong compared to $h_j$ when going slightly to the left of $j_0$, thus restoring the MBL behavior.
} and small higher-order terms, both the FM and PM regions are expected to be MBL~\cite{Huse2013}.
The approximated integrals of motion are shown in Fig.~\ref{fig:ising}c.
In the FM region, they are $\eta^z_i\eta^z_j$ for nearby $\langle ij\rangle$, and a nonlocal string $\prod_{j\leq j_0}\eta^x_j$; in the PM region, they are $\eta^x_j$.
The approximation works well at the locations away from $j_0$ by the order of $\xi$.
An energy eigenstate of the $\eta$ chain can be constructed by a simultaneous eigenstate of these integrals of motion:
\begin{equation}
    |\psi_\pm\rangle\approx\frac{1}{\sqrt{2}}\left(\bigotimes_{j\leq j_0}|a_j\rangle\pm \bigotimes_{j\leq j_0}|1-a_j\rangle\right)\otimes \bigotimes_{j> j_0}|b_j\rangle
\end{equation}
where $a_j=0,1$ and $b_j=+,-$.
In our dynamical setup where the initial state is approximately an $\eta^z$ eigenstate, the oscillation frequency $\omega_\text{sync}$ in the FM part is dominated by the level splitting between $|\psi_+\rangle$ and $|\psi_-\rangle$, corresponding to the coefficient of $\prod_{j\leq j_0}\eta^x_j$.
On the other hand, the PM region behaves essentially the same as that of the non-interacting case.
In other words, the dynamics are synchronized from the first site to around $j_0$, beyond which it becomes unsynchronized.
The larger the interaction is, the more sites get synchronized, manifestly showing that the phenomenon is interaction-driven.

To estimate $\omega_\text{sync}$, we treat $h^\text{eff}_j$ for $j\leq j_0$ as perturbations in the $\eta$ chain itself, resulting in $\omega_\text{sync}\approx 2\big|\prod_{j\leq j_0}h^\text{eff}_j/\prod_{\langle ij\rangle,j \leq j_0}J^\text{eff}_{ij}\big|$~\cite{Mbeng2024,Olund2025}.
In the simplest case where $\xi\lesssim1$ and all sites are active, this becomes $\omega_\text{sync}\approx \omega_1(0)\omega_2(0)\cdots\omega_{j_0}(0)/\Delta^{j_0-1}$, where $\omega_j(0)$ indicates the dominated frequency of site $j$ at $\Delta=0$.
However, unless $\xi\ll 1$, $\omega_j(0)$ may not be strictly increasing with $j$, making $j_0$ ill-defined. To obtain a more accurate estimation, we sort the frequencies by letting $\tilde\omega_j$ be the $j$th smallest $\omega_j(0)$, and then estimate $j_0$ and $\omega_\text{sync}$ by
\begin{equation}\label{eq:omega_sync}
    \omega_\text{sync}=\frac{\tilde\omega_1\tilde\omega_2\cdots\tilde\omega_{j_0}}{\Delta^{j_0-1}},\text{ where }\tilde\omega_{j_0}\leq\Delta<\tilde\omega_{j_0+1}.
\end{equation}
This should be a good approximation for $\Delta\ll 1$, where perturbation theory works, and the higher order terms of the $\eta$ chain are negligible.

The synchronization in the FM region is due to spontaneous symmetry breaking of the Ising symmetry $\prod_j\eta_j^x$, which comes from the mirror symmetry of the original chain.
Thus, a natural order parameter to characterize the synchronization transition is the correlator $\langle\eta^z_i\eta^z_j\rangle$, whose long-time average for a $\eta^z$ basis state approaches a finite value if $\eta_i$ and $\eta_j$ are coherently synchronized, and zero otherwise. As only $\sigma_i$ are the direct observables, one can approximately use $\langle\sigma^z_i\sigma^z_j\rangle$ as the order parameter for $\xi\lesssim 1$. A related concept has been discussed in Ref.~\cite{thomas2018quantum}.

Even in the regime where $\Delta$ cannot be treated perturbatively, synchronization is expected to persist.
To see this, let the unperturbed Hamiltonian be the full $H_0+PH_0P$ with only $H_{\rm cp}$ as a perturbation. Assuming no accidental degeneracies, the unperturbed eigenspaces with some active sites are two-fold degenerate.
The degeneracy is then lifted by the off-diagonal element $t$, and the perturbed eigenstates become a pair of cat states, implying synchronization with $\omega_\text{sync}\approx |2t|$ (note that the sites very near the center bond may still be unsynchronized since $H_{\rm cp}$ cannot be treated as a perturbation at the center).

\textit{Numerical results.} -- To verify the theory and illustrate the generic synchronized oscillations in our model, we perform numerical calculations on model (\ref{eq:model}). First, to compute the oscillating frequencies as a function of interaction, we compute the power spectrum of the autocorrelation function of $ \langle \sigma_i^z(t) \sigma_i^z(0)\rangle$ for a specific initial product state $|\psi_0\rangle$
\begin{equation}
    \mathcal S_i (\omega) = 2 \pi s_i^z \sum_{mn} c_m c_n^* (\sigma_i^z)_{mn} \delta(\omega + E_m -E_n)
    \label{eq:spectrum}
\end{equation}
where $\sigma_i^z(0)|\psi_0\rangle = s_i^z|\psi_0\rangle, c_n = \langle\psi_0|\phi_n\rangle, (\sigma_i^z)_{mn} = \langle \phi_m | \sigma_i^z |\phi_n\rangle$, and $|\phi_n\rangle$ is the many-body eigenstate.  
Since $\mathcal S_i (\omega)$ is sharply peaked for a specific disorder realization in the localized regime, we choose the median frequency as the oscillating frequency of spin $i$. 

\begin{figure}
\centering
\includegraphics[width=\linewidth]{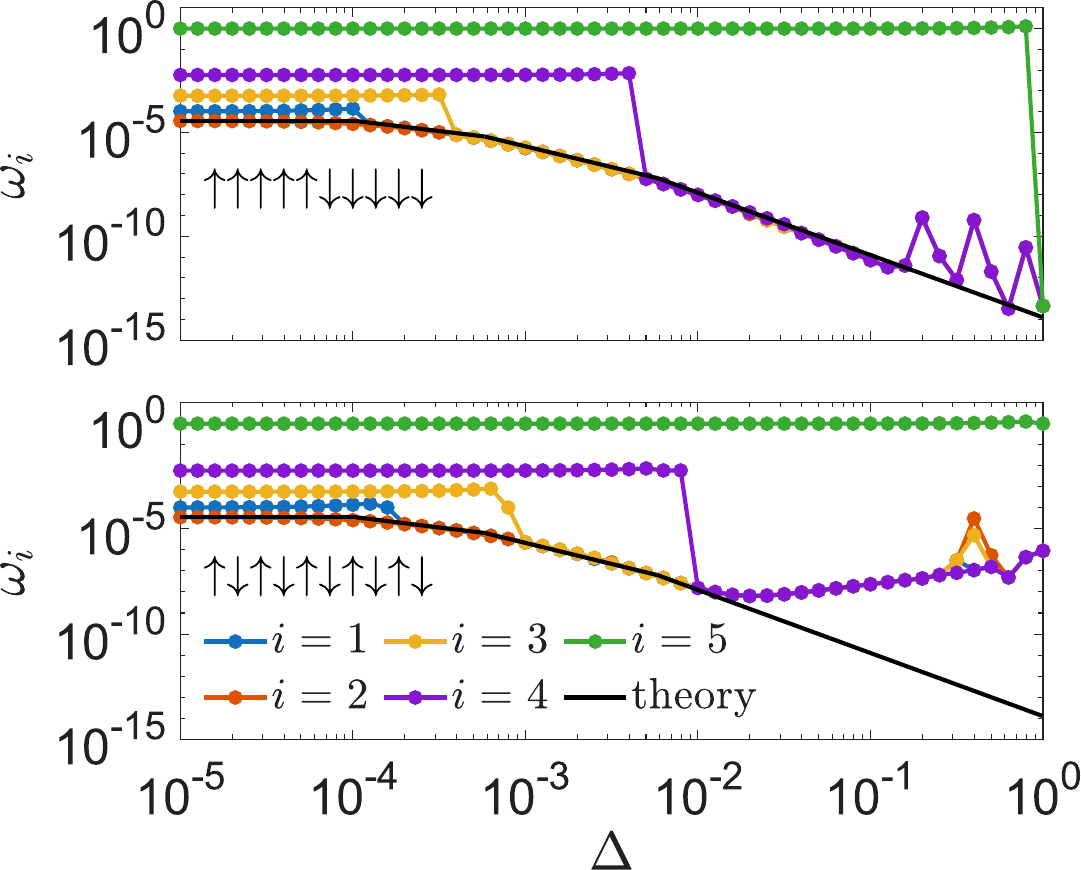}

\caption{\justifying {Median frequency of each particle with the given initial $\sigma^z$ eigenstate as a function of $\Delta$. Note that more and more particle becomes synchronized as $\Delta$ grows. The theory line for $\omega_\text{sync}$ is calculated from Eq.~(\ref{eq:omega_sync}), which works for $\Delta\ll 1$. For both figures, we consider a single disorder realization at $W=4,L=10$.}}
\label{fig:freq}
\end{figure}

Fig.~\ref{fig:freq} shows the numerical results of all $\omega_i(\Delta)$ for two initial states. The sudden coalescence of the spin frequencies at some $\Delta$ into one frequency implies the synchronization between different spins, and the threshold interaction is $\Delta \approx \omega_i$ as predicted.
The small-$\Delta$ behavior is largely independent of the initial states, and the theoretical value of the synchronized frequency in Eq.~\ref{eq:omega_sync} agrees with the numerics remarkably well, implying that the system at $W=4$ is localized enough for the theory to hold, although $W$ is only slightly larger than the reported critical value $W_c \sim 3.5$.
The value of $\Delta$ where the theory becomes inaccurate, and the synchronized frequency beyond this value, are dependent on the initial state, but the spins are still mostly synchronized. 
The fact that the spin at the midpoint of the chain may be unsynchronized is also expected, since $H_{\rm cp}$ cannot be treated as a perturbation. 
Note that no disorder averaging is needed, and the theory holds for different disorders and initial state realizations as long as the disorder strength is large, as we tested numerically. 

The other direct numerical verification, as we point out in the theory, is the infinite time average of the spin-spin correlation
\begin{equation}
   \overline C_{ij} = \lim_{T\rightarrow \infty}\frac{1}{T}\int_0^T  \langle \psi_0|e^{i H t}\sigma_i^z \sigma_j^z e^{-i H t} |\psi_0\rangle \mathrm{d}t,
\end{equation}
which is also a synchronization measure. For two unsynchronized spins, the direction of spins is uncorrelated, so that the correlation is $\overline C_{ij} \rightarrow 0$; for two completely synchronized spins, they are either aligned or anti-aligned, thus the correlation is $\overline C_{ij} \rightarrow \pm 1$.

\begin{figure}
\centering
\includegraphics[width=\linewidth]{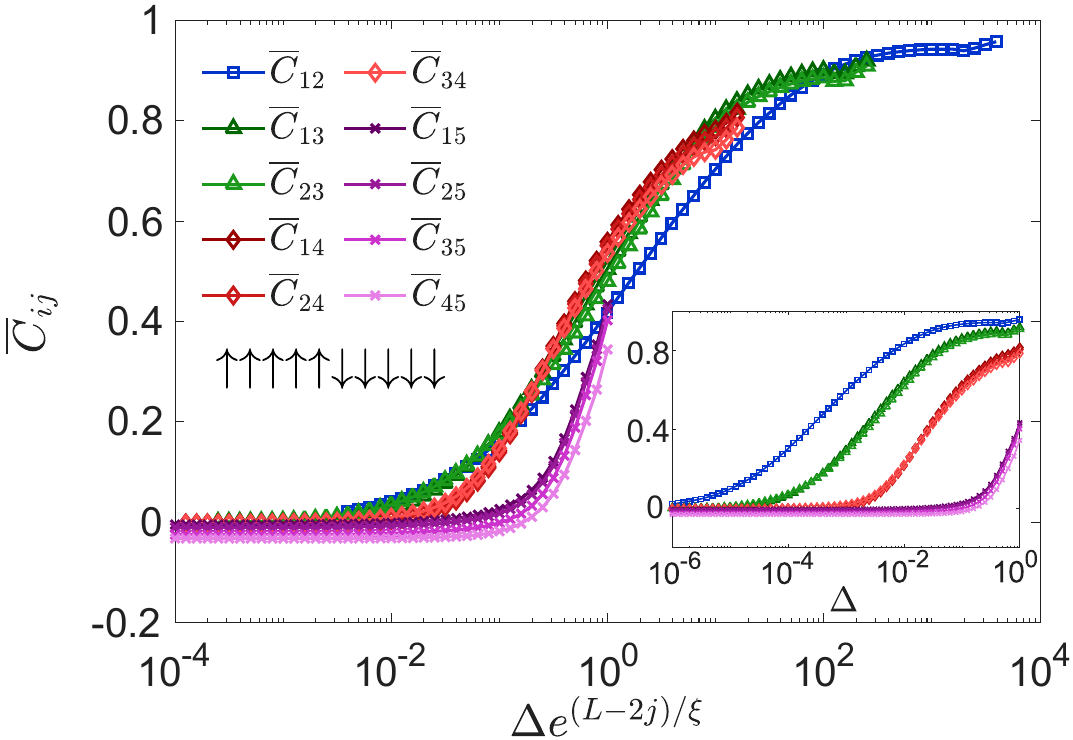}

\caption{\justifying {Infinite time averaged spin-spin correlation as a function of interaction. Parameters are $L=10,W=4,\xi=0.725$. $N=5$ particles are initially positioned at the left side of the chain.}}
\label{fig:correlation}
\end{figure}

Fig.~\ref{fig:correlation} shows all the correlations of spins on the left part of the chain as a function of interaction for a specific initial state, computed by averaging $10^6$ disorder realizations. 
In the inset, we show that the synchronized interaction only relates to the spin index $j$, the index of the spin closer to the middle of the chain, and is irrelevant to $i$. 
Moreover, it scales as $e^{-(L-2j)/\xi}$, which is exactly the scaling factor of the Ising transition as shown in Eq.~\ref{eq:critical}. 
The localization length $\xi$ is chosen as the non-interacting localization length, since the synchronized interaction is small and the localization length is largely unchanged for most of the spins. 
The slight discrepancy of the correlation curves $\bar C_{i5}$ from other curves could be due to the drift of the localization length at large interactions, or the decreasing alignment between the $\tau_i^z$ and $\sigma_i^z$. We have checked for other initial spin configurations, and the results are qualitatively the same.

\textit{Discussion and Conclusion} -- Due to experimental imperfections, one expects the mirror symmetry of the on-site potential to be explicitly broken by a small asymmetry (see Supplemental Material~\cite{SM} for details). 
Roughly speaking, this corresponds to having a minimal frequency $\omega_\text{min}$ such that if the oscillation frequency is below it, the even/odd cat states break into two LIOM configurations, and the particles participating in the oscillation freeze in their initial sites. This is similar to having a damped oscillator where the oscillation frequency is cut off from below by the damping. 
Note that even if $\omega_\text{sync}<\omega_\text{min}$ at the small $\Delta$, one may still have $\omega_\text{sync}>\omega_\text{min}$ near $\Delta\sim 1$ for some initial states due to the non-monotonicity of $\omega_\text{sync}$ (as in the second panel of Fig.~\ref{fig:freq}). Therefore, we expect the synchronized oscillation of many particles to be most easily experimentally observable in this regime. 

In the perturbative regime, according to Eq.~\ref{eq:omega_sync}, the synchronized frequencies approximately scales as $\omega_{\rm sync} \sim e^{-L^2/ 2\xi}$ at fixed interaction and disorder strengths. Thus, the synchronized oscillations become fragile to external perturbations at large system sizes. 
In the regime where $\Delta \sim 1$, the scaling of synchronization frequency with the system size depends on the initial states in a complicated way. 
A detailed analysis of this regime is beyond the scope of this paper and could be the subject of future work.
Regardless of these complications, the synchronization phenomenon should be more easily observed in small systems.

It is worth pointing out the similarity between the synchronized behaviors in our system and the measure synchronization in Ref.~\cite{hampton1999measure}: 1) Both systems are conservative. 2) Both systems have a synchronization transition at a critical parameter. 3) In the synchronized regime, both systems are not completely synchronized (compared to the limit-cycle synchronization), but the phase difference of the oscillations is bounded. (In our model, the phase of the oscillations can be considered as the $\phi_i(t)\equiv \arccos \langle \sigma_i^z(t) \rangle$, with $\phi_i(t)$ being continuous.) These similarities imply that one may also characterize the synchronization in quantum systems through energy eigenstates, in a way analogous to the measure of classical orbits.

We have shown the coherently synchronized oscillations of particles in an MBL system with mirror symmetry. An effective Ising chain has been derived based on the LIOMs, and the synchronization can be interpreted as a spontaneous symmetry breaking of the Ising chain. The analytic form of synchronized oscillating frequency has been obtained, which agrees well with the numerical results at small interaction. The interaction for the appearance of synchronization has been estimated from the theory and verified by the numerics.

\textit{Acknowledgement} -- The authors thank D.\ Vu for useful discussions.
This work is supported by the Laboratory for Physical Sciences.
The authors acknowledge the University of Maryland supercomputing resources (\href{https://hpcc.umd.edu}{https://hpcc.umd.edu}) made available for conducting the research reported in this paper.

\bibliographystyle{apsrev4-1}
\bibliography{ref}

\clearpage
\renewcommand\thefigure{S\arabic{figure}}    
\setcounter{figure}{0} 
\renewcommand{\theequation}{S\arabic{equation}}
\setcounter{equation}{0}
\renewcommand{\thesubsection}{SM\arabic{subsection}}

\onecolumngrid
\vspace{1em}
\begin{center}
{\large \textbf{Supplemental Material to ``Coherently synchronized oscillations in many-body localization"}}
\end{center}
\vspace{1em}

\twocolumngrid
\section{Effects of imperfection}
We mention that a small asymmetry may break the cat states into LIOM configurations. Phenomenologically, when the perturbation increases, the dynamics of spins transform from collectively oscillating to non-oscillating, i.e., frozen in their initial states. 
The magnitude of the perturbation inducing such a transition is related to the synchronized frequencies $\omega_{\rm sync}$. To show this, we add a small onsite asymmetric disorder to the Hamiltonian
\begin{equation}
    H' = \frac{\epsilon }{2}\sum_{i=1}^{L} h_i \sigma_i^z
\end{equation}
where $h_i$ is sampled from the same interval $[-W,W]$, $\epsilon$ is the parameter that controls the magnitude of the perturbation. 

\begin{figure}
  \centering
  \includegraphics[width=1\linewidth]{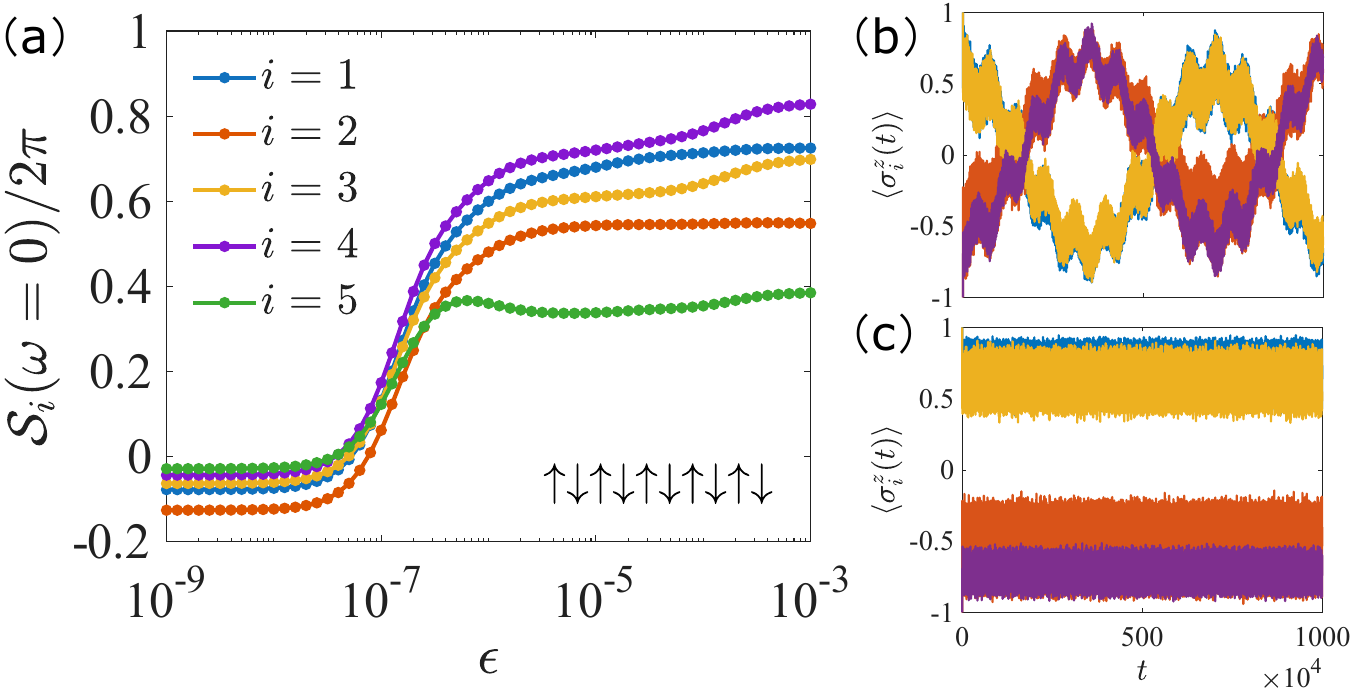}
  \caption{\justifying{Spin dynamics in the presence of a small asymmetric disorder. (a) Zero-frequency components of the power spectrum as a function of $\epsilon$. (b) The spin dynamics is collectively synchronized oscillating without asymmetry ($\epsilon = 0$), and is (c) collectively ``frozen" with a relatively large asymmtry $\epsilon = 10^{-5} > \omega_{\rm sync}$.  For all figures, the disorder realization is the same as in Fig.~{\red 3} in the main text, other parameters are $\Delta = 1,L=10$, and the initial state is a N\'eel state.}}
  \label{fig:perb}
\end{figure}

We compute the amplitudes of non-oscillating modes, using the zero-frequency components of the power spectrum, as a function of $\epsilon$. Fig.~\ref{fig:perb}(a) shows that when $\epsilon \gtrsim \omega_{\rm sync}$, the non-oscillating modes dominate the synchronized oscillating modes. The transition of local spin dynamics is shown in Fig.~\ref{fig:perb}(b) and (c), where the spin transforms from the synchronized oscillating mode to the non-oscillating (``frozen") mode.

\end{document}